\documentclass[11pt, A4]{article}

\usepackage{graphicx}
\usepackage{epstopdf}
\usepackage{amsmath}
\usepackage{amssymb}
\usepackage{amsfonts}
\usepackage{amsthm}
\usepackage[usenames]{color}
\usepackage{array}

\usepackage[
      colorlinks=true,
      linkcolor=blue,
      urlcolor=blue,
      filecolor=blue,
      citecolor=red,
      pdfstartview=FitV,
      pdftitle={},
      pdfauthor={},
      pdfsubject={},
      pdfkeywords={},
      pdfpagemode=None,
      bookmarksopen=true
]{hyperref}

\usepackage{epsfig}

\usepackage{hyperref}

\def\beq{\begin{eqnarray}}
\def\eeq{\end{eqnarray}}

\def \RR{{\mathbb{R}}}

\def\mf{\mathfrak}
\def\be{\begin{equation}}
\def\ee{\end{equation}}
\def\bea{\begin{eqnarray}}
\def\eea{\end{eqnarray}}

\newcommand{\rom}[1]{\mathrm{#1}}

\def\cA{\mathcal{A}}

\def\cL{\mathcal{L}}
\def\cM{\mathcal{M}}

\def\cV{\mathcal{V}}

\def\mf{\mathfrak}
\def\nn{\nonumber}

\numberwithin{equation}{section}

\begin{document}

\begin{centering}

  \vspace{0cm}

\textbf{\Large{
Subtracted Geometry from Harrison Transformations: II}}

 \vspace{0.8cm}

  {\large Anurag Sahay and Amitabh Virmani}

  \vspace{0.5cm}

\begin{minipage}{.9\textwidth}\small  \begin{center}
Institute of Physics, Sachivalaya Marg, Bhubaneswar, Odisha, India 751005 \\
  \vspace{0.5cm}
{\tt anurag, virmani@iopb.res.in}
\\ $ \, $ \\

\end{center}
\end{minipage}

\end{centering}

\begin{abstract}
We extend our previous study (arXiv:1203.5088) to the case of five-dimensional multi-charge black holes, thus showing that these configurations and their subtracted geometries also lie in a 3d duality orbit. In order to explore the 3d duality orbit, we do a timelike reduction from 5d to 4d and a spacelike reduction from 4d to 3d. We present our analysis in the notation of Euclidean N=2 supergravity and its c-map. We also relate our analysis to that of Cveti\v{c}, Guica, and Saleem.
\end{abstract}

%\newpage
\tableofcontents
\newpage

\setcounter{equation}{0}
\section{Introduction and Summary}
String theory has come a long way in addressing the question of entropy for extremal black holes. However,  very
little is known about microscopic origins of  thermodynamic properties of general non-extreme
black holes. An obvious obstacle one faces in making progress on this problem is that
the specific heats of non-extreme black holes in asymptotically flat settings are typically negative,
whereas those of unitary field theories are typically positive. To get around some of these difficulties
Cveti\v{c} and Larsen \cite{Cvetic:2011hp, Cvetic:2011dn}  have put forward the
proposal of subtracted geometry.

The notion of subtracted geometry offers several attractive features. Firstly, it makes the idea of
hidden conformal symmetry of Castro, Maloney, and Strominger \cite{Castro:2010fd}
geometrical. Secondly, it provides us with a classical geometry that can be regarded, in some respect, as
the ``near-horizon'' analog for non-extreme black holes.
Thirdly, perhaps the most attractive feature of  subtracted geometries is that they lift to AdS$_3$ times a
sphere in one higher dimension. These features have attracted the attention of several researchers \cite{
deBoer:2011zt,
Cvetic:2012tr,
Baggio:2012db,
Chakraborty:2012nu}.
However, it remains to be seen how these ideas pan out in paving a path towards
a detailed microscopic understanding of
general non-extreme black holes in string theory. A summary of the current literature on this subject is as follows.

Since the beginning an important aspect in the discussion of subtracted geometry has been  how to obtain these geometries
systematically starting from a black hole solution. The initial papers \cite{Cvetic:2011hp, Cvetic:2011dn} focused on the STU model
where the subtraction procedure was implemented as an adjustment of the
warp factor of multi-charge asymptotically flat black holes. It was then not clear how to implement such a procedure for other
settings. Around that time, Bertini, Cacciatori, and Klemm \cite{Bertini:2011ga} realized that $\mathrm{SU}(2,1)$ Harrison transformations\footnote{These transformations are
also sometimes known as Kinnersley-Chitre transformations.} of
Einstein Maxwell theory, which map 4d
Schwarzschild solution to AdS$_2 \times \mathrm{S}^2$, also map Schwarzschild hidden conformal symmetry generators
to isometries of AdS$_2$. Significant progress happened when
Cveti\v{c} and Gibbons \cite{Cvetic:2012tr} showed that the subtracted geometry
can be obtained by a scaling limit of asymptotically flat multi-charge black holes. They also conjectured that these geometries
 lie in a three-dimensional duality orbit of the original black hole, and hence should be obtainable using
a particular Harrison transformation on the original black holes. This conjecture was confirmed in our previous work \cite{Virmani}. For earlier
related work see \cite{Boonstra:1997dy, Sfetsos:1997xs}.
Motivated by these developments Baggio, de Boer, Jottar, and Mayerson \cite{Baggio:2012db} argued that at least in the 4d static case the
adjustment of the warp factor can be implemented dynamically by means of an interpolating flow. They also constructed an interpolating solution.
From their point of view a dual
description of the
5d uplift of the 4d multi-charge asymptotically flat black hole can be obtained by turning on certain specific irrelevant deformations in the Maldacena-Strominger-Witten (MSW) CFT.
In another line of investigation Chakraborty, Jana, and Krishnan \cite{Chakraborty:2012nu} have studied various aspects of
interpolating solutions in the extremal limit and their relation to the attractor
mechanism. They have also proposed a more general notion of subtracted geometries. Recently Cveti\v{c}, Guica, and Saleem \cite{Cvetic:2013cja}
 have constructed interpolating solution for both 4d and 5d rotating black holes. For other related developments see also \cite{rel1, rel2, rel3}.

In this paper we extend our previous study \cite{Virmani}
to the case of 5d multi-charge black holes. Our study provides a somewhat different perspective on the five-dimensional analysis of \cite{Cvetic:2013cja}.
We show that 5d multi-charge, rotating, non-extreme black hole configurations and their subtracted geometries
lie in a 3d duality orbit. The duality orbit we explore is of a rather special type: we first
do a timelike reduction (for details on timelike reduction see \cite{Hull:1998br, Cremmer:1998em})  from 5d to 4d and then a spacelike reduction from 4d to 3d. Since we do timelike reduction first,  we get
a Euclidean theory in 4d. This theory is nothing but the Euclidean version
of the N=2 STU supergravity.
We make connection with the well developed formalism of Euclidean N=2 supergravity \cite{Mohaupt1, Mohaupt2, Mohaupt3, Gutowski:2012yb}.

The rest of the paper is organized as follows. In section \ref{sec:dimred} we perform the	
dimensional reduction.  In section \ref{sec:5dsub} we present
the main result of our investigation, first for static black holes and then for rotating black holes.
 In this section we also relate our analysis to that of Cveti\v{c}, Guica, and Saleem \cite{Cvetic:2013cja}.

\section{Dimensional Reduction}
\label{sec:dimred}
In this section we present an appropriate dimensional reduction of five-dimensional U(1)$^3$ supergravity to three dimensions.
\subsection{Timelike Reduction: 5d to 4d}
Our starting point is the Lagrangian of five-dimensional U(1)$^3$ supergravity
\begin{equation}
\label{eqn:Lagrangian5d}
\mathcal{L}_5 = R_5 \star_5 \mathbf{1}  - \frac{1}{2} G_{IJ} \star_5 dh^I \wedge  dh^J
- \frac{1}{2}G_{IJ}\star_5 F^I_{[2]} \wedge F^J_{[2]}
+ \frac{1}{6} C_{IJK} F^I_{[2]} \wedge F^J_{[2]} \wedge A^K_{[1]},
\end{equation}
where $C_{IJK} = |\epsilon_{IJK}|$, with $I=1,2,3$, and $G_{IJ}$ is diagonal with entries $G_{II}
= (h^I)^{-2}$. The scalars $h^I$s satisfy the constraint $h^1 h^2 h^3 = 1$ that must be solved before computing variations of the action in order
to obtain equations of motion for various fields.

To perform KK reduction we parameterize our 5d spacetime as
\be
\label{an1}
ds^2 = \epsilon_1 f^2  (dt + \check A^0_{[1]})^2 + f^{-1}ds^2_4,
\ee
and 5d vectors as
\be
\label{an2}
A^I_{[1]} = \chi^{I}(dt + \check A^0_{[1]}) + \check A^I_{[1]},
\ee
where we use $\epsilon_1$ to keep track of minus signs. When $\epsilon_1$ is $+1$ we are performing the standard spacelike reduction that, for example, is presented in detail in \cite{Virmani}. The case of interest for the present discussion is $\epsilon_1 = -1$.

Upon KK reduction the 4d graviphoton $\check A^0_{[1]}$ and the 4d vectors $\check A^I_{[1]}$ together form a symplectic vector $\check A^\Lambda_{[1]}$ with $\Lambda = 0,1,2,3$. We define the field strength for the symplectic vector to be simply $F_{[2]}^\Lambda = d \check A_{[1]}^\Lambda$.

From the Lagrangian \eqref{eqn:Lagrangian5d} using the ansatzes \eqref{an1} and \eqref{an2} we obtain,
\bea
\mathcal{L}_4 &=&  R \star_4 \mathbf{1} - \frac{1}{2}G_{IJ} \star_4 d h^I \wedge
d h^J  - \frac{3}{2f^2}\star_4 df \wedge df   - \epsilon_1 \frac{f^3}{2}\star_4
\check F_{[2]}^0 \wedge \check F_{[2]}^0 \label{STUaction} \\
&& - \epsilon_1 \frac{1}{2 f^2}G_{IJ} \star_4 d \chi^I \wedge d \chi^J  - \frac{f}{2}
G_{IJ} \star_4 (\check F_{[2]}^I +\chi^I \check F_{[2]}^0) \wedge (\check
F_{[2]}^J +\chi^J \check F_{[2]}^0) \nn
\\
&&+\frac{1}{2}C_{IJK}\chi^I \check F_{[2]}^J  \wedge \check F_{[2]}^K  +
\frac{1}{2} C_{IJK}\chi^I \chi^J \check F_{[2]}^0  \wedge \check F_{[2]}^K + \frac{1}{6}
C_{IJK} \chi^I \chi^J \chi^K \check F_{[2]}^0 \wedge \check F_{[2]}^0\, .
 \nn
\eea
Note that there are two signs in the above expression, namely the kinetic term for the graviphoton $\check F_{[2]}^0$ and the kinetic terms for the scalars $\chi^I$s, that depend on the sign $\epsilon_1$.

\subsection{Euclidean Supergravity}
The reduced Lagrangian \eqref{STUaction} can also be obtained using prepotential formalism
\cite{Ceresole:1995jg}
 of  Euclidean $N=2$ supergravity
\cite{Mohaupt1, Mohaupt2, Mohaupt3, Gutowski:2012yb}. We briefly summarize this formalism.  It uses the so-called split complex numbers. Recall that split complex numbers \cite{split} satisfy the standard conjugation relation but the imaginary unit $e$ squares to $+1$ instead of $-1$,
\be
 \bar{e} = - e, \qquad \qquad e^2 = +1.
\ee
As in the Lorentzian case, the action of Euclidean $N=2$ supergravity coupled to $n$ vector multiplets is also governed by a prepotential function $F$, which is now a function of $n+1$ split complex variables $X^\Lambda$  $(0 \le \Lambda \le n)$.  The gauge invariant bosonic degrees of freedom are the Euclidean metric $g_{\mu \nu}$, split complex scalars $X^\Lambda$, and $n+1$ vectors $\check A^\Sigma$. The Lagrangian \cite{Ceresole:1995jg} is given by \cite{Mohaupt1, Mohaupt2, Mohaupt3, Gutowski:2012yb}
\be
{\cal L}_4 = R \star_4 \mathbf{1} - 2 g_{I \bar{J}} \star_4 d X^I \wedge d \bar{X}^{\bar{J}} + \frac{1}{2} \check F^\Lambda_{[2]} \wedge \check G_{\Lambda [2]},
\ee
where $\check F^\Lambda_{[2]} = d \check A^\Lambda_{[1]}$. The indices $I, J$ run from 1 to $n$, and $g_{I \bar{J}} = \partial_I \partial_{\bar{J}} K$ with the potential
\be
K = - \log \left[ - e (\bar{X}^\Lambda F_\Lambda - \bar{F}_\Lambda X^\Lambda)\right],
\ee
and where $F_\Lambda = \partial_{\Lambda} F$. The two form  $\check G_{\Lambda [2]}$ is defined
as
\be
\check G_{\Lambda [2]} = (\mathrm{Re} N)_{\Lambda \Sigma} \check F^{\Sigma}_{[2]} + (\mathrm{Im} N)_{\Lambda \Sigma} \star_4 \check F^{\Sigma}_{[2]},
\ee
where the split complex symmetric matrix $N_{\Lambda \Sigma}$ is constructed from the prepotential as
\be
N_{\Lambda \Sigma} = \bar{F}_{\Lambda \Sigma} + 2 e \frac{(\mathrm{Im} F \cdot X)_\Lambda (\mathrm{Im} F \cdot X)_\Sigma}{X \cdot\mathrm{Im} F \cdot X },
\ee
and where $F_{\Lambda \Sigma} = \partial_\Lambda \partial_\Sigma F$. Most of the above formulas are similar to the ones used in our previous work \cite{Virmani}. However, note that there are some sign changes with respect to \cite{Virmani}, and most importantly the imaginary unit $i$  of the standard complex numbers is replaced with the imaginary unit $e$ of the split complex numbers at all places.

We are interested in the Euclidean STU model. For this model $n=3$ and the prepotential function takes the form
\be
F(X) = - \frac{X^1 X^2 X^3}{X^0}.
\ee
Using the gauge fixing $X^0 = 1$ and $X^I = - \chi^I + e f h^I$,
the resulting Lagrangian can be shown to be identical to \eqref{STUaction} with $\epsilon_1 = -1$.
\subsection{Spacelike Reduction: 4d to 3d}
Now we perform a spacelike reduction from four to three dimensions to obtain an
$$\mathrm{SO}(4,4)/(\mathrm{SO}(2,2) \times \mathrm{SO}(2,2))$$
coset model. The denominator subgroup is a different $\mathrm{SO}(2,2) \times \mathrm{SO}(2,2)$ compared to the one used in \cite{Virmani} where the reduction was first done over a spacelike direction and then over a timelike direction.

For this KK reduction of the Lagrangian \eqref{STUaction} we parameterize our four-dimensional Euclidean space as
\be
ds^2_4 = e^{2U} (dz + \omega_3)^2 + e^{-2 U}ds^2_3,
\ee
and the 4d one-forms as
\be
\check A^\Lambda_{[1]} = \zeta^\Lambda (dz + \omega_3) + A_3^\Lambda,
\ee
where $A_3^\Lambda$ and $\omega_3$ are three-dimensional one-forms. Since now the reduction is done differently, the dualization equations are different compared to \cite{Virmani}. To find the correct dualization equations we proceed as in the lecture notes by Pope \cite{Pope}.

For both $A_3^\Lambda$ and $\omega_3$ we define the field strengths simply as $F_3^\Lambda := d A_3^\Lambda$ and $F_3 := d \omega_3$. The procedure of dualization interchanges the role of Bianchi identities and the field equations. The easiest way to achieve dualization is to treat $F_3^\Lambda$ and $F_3$ as fundamental fields in their own right. We impose the Bianchi identities by adding the following Lagrange multiplier terms to the 3d Lagrangian
\be
- \tilde \zeta_\Lambda F_3^\Lambda - \frac{1}{2}(\sigma + \zeta^\Lambda \tilde \zeta_\Lambda) F_3.
\ee
Thus the total three dimensional Lagrangian we consider is
\bea
{\cal L}_3' &=& R \star_3 \mathbf{1} - 2 \star_3 d U \wedge dU - \frac{1}{2} e^{4U} \star_3 F_3 \wedge F_3  - 2 g_{I\bar{J}} \star_3 d X^{I} \wedge d \bar{X}^{\bar{J}} \nn \\
 & & + \frac{1}{2} e^{2U}(\mathrm{Im}N)_{\Lambda \Sigma} \star_3 (F_3^\Lambda + \zeta^\Lambda F_3)\wedge (F_3^\Sigma + \zeta^\Sigma F_3) + \frac{1}{2} e^{-2U} (\mathrm{Im}N)_{\Lambda \Sigma}\star_3 d \zeta^\Lambda \wedge d \zeta^\Sigma \nn \\
 & & + (\mathrm{Re}N)_{\Lambda \Sigma} \: d \zeta^\Lambda \wedge (F_3^\Sigma + \zeta^\Sigma F_3)
 - \tilde \zeta_\Lambda d F_3^\Lambda - \frac{1}{2}(\sigma + \zeta^\Lambda \tilde \zeta_\Lambda) d F_3.
 \label{Lag3d1}
\eea
Clearly, variations of this Lagrangian with respect to $\sigma$ and $\tilde \zeta_\Lambda$ give the required Bianchi identities.  Equation for $F_3^\Sigma$ and $F_3$ are purely algebraic. These equations allow us to do the dualizations of the one-forms. We find
\be
\label{dual1}
- d \tilde \zeta_\Lambda = e^{2U}(\mathrm{Im}N)_{\Lambda \Sigma} \star_3 (F_3^\Sigma + \zeta^\Sigma F_3) + (\mathrm{Re}N)_{\Lambda \Sigma} d\zeta^\Sigma,
\ee
and
\be
\label{dual2}
- d \sigma = - 2 e^{4U} \star_3 F_3 + \tilde \zeta_\Lambda d \zeta^\Lambda - \zeta^\Lambda d \tilde\zeta_\Lambda.
\ee
Substituting these back into Lagrangian \eqref{Lag3d1} we find  ${\cal L}_3'$ takes the form (from now on we drop the prime)
\bea
\cL_3 = R \star_3 \mathbf{1} -\frac{1}{2} G_{ab} \partial \varphi^a \partial
\varphi^b,
\eea
 where the target space is a  Lorentzian manifold (as expected). It is parameterized by 16 scalars
 $\varphi^a$ and is of signature $(8,8)$. The metric in our conventions
 is
\bea
 G_{ab}d\varphi^a d\varphi^b &=& 4 dU^2 + 4 g_{I \bar{J}}dz^I d\bar{z}^{\bar J} -
 \frac{1}{4}e^{-4U} \left( d\sigma +  \tilde \zeta_\Lambda d \zeta^\Lambda -
 \zeta^\Lambda d \tilde \zeta_\Lambda \right)^2 \label{cmapst} \\
 && \hspace{-2cm} + e^{-2U}\left[ -(\mbox{Im} N)_{\Lambda \Sigma}d\zeta^\Lambda
 d\zeta^\Sigma + ((\mbox{Im} N)^{-1})^{\Lambda \Sigma} \left( d\tilde
 \zeta_\Lambda +(\mbox{Re} N)_{\Lambda \Xi} d\zeta^\Xi \right)  \left( d\tilde
 \zeta_\Sigma +(\mbox{Re} N)_{\Sigma \Gamma} d\zeta^\Gamma \right) \nonumber \right].
\eea
It is a different analytic continuation of the
 c-map of Ferrara and Sabharwal  \cite{Ferrara:1989ik}\footnote{Extra care must be exercised in checking this analytic continuation.
Recall that in all the expressions above, real and imaginary parts refer to the split complex numbers, whereas in Ferrara and Sabharwal
 \cite{Ferrara:1989ik} or in \cite{Bossard:2009we} the real and imaginary parts refer to the standard complex numbers. Perhaps the easiest way to check this analytic continuation is to express the three-dimensional Lagrangian explicitly in terms of various scalars and then compare it with the analytically continued expressions of Ferrara and Sabharwal.} compared to the one used in \cite{Virmani}.
 The analytic continuation is as follows\footnote{In these equations $i$ refers to the imaginary unit of the standard complex numbers.}
\bea
\chi^I &\to& i \chi^I \\
\sigma &\to& i \sigma \\
\zeta^0 &\to& - i \zeta^0 \\
\tilde \zeta_0 &\to& - \tilde \zeta_0 \\
\zeta^I &\to&   \zeta^I \\
\tilde \zeta_I &\to& i \tilde \zeta_I.
\eea
Note that only 8 scalars pick up a factor of $i$, so that the signature of the target space is (8,8).

The symmetric space \eqref{cmapst} can  be parameterized in the Iwasawa gauge by the coset
element \cite{Bossard:2009we}
\be
\label{iwasawa}
\cV = e^{- U \, H_0} \cdot \prod_{I=1,2,3}\left(e^{-\frac{1}{2} \left[\log (f h^I)\right] H_I} \cdot e^{ \chi^I E_I} \right) \cdot
e^{-\zeta^\Lambda E_{q_\Lambda}-  \tilde \zeta_\Lambda E_{p^\Lambda}}\cdot
e^{-\frac{1}{2}\sigma E_0}.
 \ee
For the Lie algebra generators we use the same notation as \cite{Virmani} (see also appendix \ref{app:so44}). The metric \eqref{cmapst} is obtained through the Maurer-Cartan one-form $\theta = d \cV \cdot \cV^{-1}$ as follows (for details see for example \cite{Pope,G21})
\bea
G_{ab} d \varphi^a d \varphi^b &=& \mathrm{Tr}(P_\star P_\star), \\
P_\star &=& \frac{1}{2} \left(\theta - \tilde \tau(\theta) \right),
\eea
where the involution $\tilde \tau$ that defines the coset is:
\begin{align}
&\tilde\tau(H_0) = -H_0, %\qquad \qquad
& & \tilde\tau(H_I) = - H_I, \\
&\tilde\tau(E_0) = + F_0, %&\qquad \qquad&
& &\tilde\tau(E_I) = + F_I, \\
&\tilde\tau(E_{q_{0}}) = + F_{q_{0}}, %\qquad \qquad
& &\tilde\tau(E_{q_{I}}) = - F_{q_{I}}, \\
&\tilde\tau(E_{p^{0}}) = - F_{p^{0}}, %\qquad \qquad
& &\tilde\tau(E_{p^{I}}) = + F_{p^{I}}.
\end{align}

\section{5d Charged Black Holes and their Subtracted Geometry}
\label{sec:5dsub}
In this section we make use of the formalism presented in the previous section and show that subtracted geometry of five-dimensional
three-charge black hole lies in a three-dimensional duality orbit of the black hole itself.  Cveti\v{c}, Guica, and Saleem have recently obtained \cite{Cvetic:2013cja}
a closely related result. They have shown that in the five-dimensional case subtracted geometry can be obtained from STU
transformations of the Euclidean STU supergravity. We also obtain this result, however, our presentation and analysis
is complementary to theirs. We show this by establishing that the three-dimensional duality transformations that we apply to obtain subtracted geometry
are  in fact part of the four-dimensional duality group. We establish this using the analysis of Bossard, Nicolai, and
Stelle \cite{Bossard:2009at}.

The reason we work with three-dimensional duality orbits is manifold. (i) Once we realized that from 5d to 4d timelike reduction
is required to perform our analysis we found it natural to relate our analysis to the well developed Euclidean N=2 supergravity formalism
\cite{Mohaupt1, Mohaupt2, Mohaupt3, Gutowski:2012yb} including its c-map. (ii) The original construction of five-dimensional three-charge black
hole by Cveti\v{c} and Youm \cite{Cvetic:1996xz} was done using 3d duality transformations. We found it useful to relate our analysis to theirs at certain
intermediate steps. (iii) Lastly, four-dimensional Euclidean STU transformations are part of the three-dimensional duality transformations. Thus, our
analysis offers a different perspective on the results of Cveti\v{c}, Guica, and Saleem.

\subsection{Three-dimensional Duality Orbit}
To explore 3d duality orbit we proceed as in \cite{Virmani, G21}. Having constructed the coset representative $\cV$ we define the generalized
transposition $\sharp(x)= - \tilde \tau (x), \forall x \in \mathfrak{so}(4,4).$ Next we encode all 16 scalars of the
$\mathrm{SO}(4,4)/\mathrm{SO}(2,2) \times \mathrm{SO}(2,2)$ coset in a matrix $\cM$ defined as $\cM = (\cV^\sharp) \cV.$
Under $\mathrm{SO}(4,4)$ group action the matrix $\cM$ transforms as
$ \cM \to \cM' = g^{\sharp} \cM g. $
The transformed solution is constructed using the new matrix $\cM'$. Since the involution $\tilde \tau$ is different in the present case
compared to the one used in \cite{Virmani}, a separate Mathematica implementation is required for extracting scalars from the matrix $\cM$. Similarly when dualizing
appropriate scalars to one-forms additional care is required in the present case since the real and imaginary parts of the
matrix $N$ in equation \eqref{dual1}
refer to the real and imaginary parts with respect to \emph{split-complex} numbers.

\subsection{5d Subtracted Geometry from Harrison Transformations}
As in our previous work \cite{Virmani} to obtain subtracted geometry of three-charge five-dimensional black hole we act on the black hole with a series a
transformations.  This investigation was initiated in \cite{Bertini:2011ga, Cvetic:2012tr}. The key insight that our previous work \cite{Virmani}
brought to this analysis was that certain \emph{negative roots} of Lie algebra $\mathfrak{so}(4,4)$ are required to perform the appropriate
Harrison boosts. For the four-charge black hole in four-dimensions three negative roots were used. Each negative root brings
down a power of $r$ in the warp factor $\Delta$. As a result, after the application of these transformations $\Delta$ grows linearly with
$r$, whereas for asymptotically flat black holes it grows as $r^4$ \cite{Cvetic:2011dn}.

For the five-dimensional analysis very similar discussion applies. Having constructed the three-charge black hole by the action
of \cite{Cvetic:1996xz}
\be
g_\rom{charging} = \exp[ \beta_1 (E_1 + F_1)] \cdot \exp[\beta_2 (E_2 + F_2)] \cdot \exp[ \beta_3 (E_3 + F_3)]
\ee
on the five-dimensional Schwarzschild black hole,
we act with
\be
g_\rom{subtraction} = \exp [F_1 + F_2].
\label{groupaction:subtraction}
\ee
Note that in this step we exponentiate only two negative roots of the $\mathfrak{so}(4,4)$ Lie algebra. Each negative root brings down the power of $r$ by two
in the five-dimensional warp factor and as a result $\Delta$ grows as $r^2$ after the action of these generators. In the asymptotically flat space
it grows as $r^6$ \cite{Cvetic:2011hp}.

As in the 4d case  \cite{Virmani} one also needs to perform certain scaling transformations to get the subtracted geometry in the form
of \cite{Cvetic:2012tr}. These transformations are as follows
\be
g_\rom{scaling} = \exp[-c_1 H_1 - c_2 H_2 - c_3 H_3].
\ee
As the next step,  we change variables following the suggestion of \cite{Cvetic:2012tr} and choose $c_1,c_2,c_3$ in some specific way. The choice
%\begin{align}
\bea
\beta_1 = c_1 =  \frac{1}{4} \ln \left[(\Pi_c^2 - \Pi_s^2)\gamma_1\right],\qquad \qquad
\beta_2 = c_2 = \frac{1}{4} \ln \left[(\Pi_c^2 - \Pi_s^2)\gamma_2\right],
\eea
and
%c_1 &=& \frac{1}{4} \ln \left[(\Pi_c^2 - \Pi_s^2)\gamma_1 \right], \\
%c_2 &=& \frac{1}{4} \ln \left[(\Pi_c^2 - \Pi_s^2)\gamma_2 \right], \\
\be
\beta_3 = \sinh^{-1}\left[ \frac{\Pi_s}{\sqrt{\Pi_c^2 - \Pi_s^2}}\right], \qquad \qquad c_3 = - \frac{1}{2} \ln \left[(\Pi_c^2 - \Pi_s^2)\right],
\ee
%nd{align}
leads to subtracted geometry in exactly the form as given in the appendix of \cite{Cvetic:2012tr}.
The resulting geometry still has $\gamma_1$ and $\gamma_2$ are parameters, but they appear exclusively as constant terms in five-dimensional
vector fields
and hence can be gauged away.

For simplicity we first present various fields and further details for the static case.
After the action of $g_\rom{charging}$ on 5d Schwarzschild black hole
various five-dimensional fields take the following form. The metric is
\be
ds_5^2 = - \Delta^{-\frac{2}{3}} r^2(r^2 -2m) dt^2 + \Delta^{\frac{1}{3}}d \tilde s^2_4
\label{staticmetric5d}
\ee
where
\be
d \tilde s^2_4 = \frac{1}{r^2 -2 m}dr^2 +d\theta^2 + \sin^2 \theta d\phi^2 + \cos^2 \theta d\psi^2,
\ee
and where the warp factor $\Delta$ is
\be
\Delta = r^6 H_1 H_2 H_3,
\ee
with the Harmonic functions
\be
H_I = 1 + \frac{2 m \sinh^2 \beta_I}{r^2}.
\ee
The five-dimensional vectors and scalars respectively take the form
\be
A_I = \frac{m \sinh 2\beta_I}{r^2 H_I}, \qquad h_I = H_I^{-1} (H_1 H_2 H_3)^{\frac{1}{3}}.
\ee
%\be
%h_1 = (H_1^{-2} H_2 H_3)^{\frac{1}{3}}, \qquad h_2 = (H_2^{-2} H_1 H_3)^{\frac{1}{3}}, \qquad h_3 = (H_3^{-2} H_1 H_2)^{\frac{1}{3}}.
%\ee
On this solution\footnote{The KK reductions are done over first $t$ and then over $\psi$.} after the action of $g_\rom{subtraction}$ and $g_\rom{scaling}$ we get
%\be
%ds_5^2 = - \Delta^{-\frac{2}{3}} r^2(r^2 -2m) dt^2 + \Delta^{\frac{1}{3}}d \tilde s^2_4
%\ee
%with
\be
\Delta =  4 m^2 \left(r^2 (\Pi_c^2-\Pi_s^2) + 2 m \Pi_s^2\right),
\ee
in the metric \eqref{staticmetric5d} and $d\tilde s_4^2$ remains unchanged. The new 5d vectors are scalars are respectively
\bea
A^1 &=& \frac{-r^2 + m + m \sqrt{\gamma_1 (\Pi_c^2 - \Pi_s^2)}}{2m} dt \equiv -\frac{r^2}{2m}dt, \\
A^2 &=& \frac{-r^2 + m + m \sqrt{\gamma_2 (\Pi_c^2 - \Pi_s^2)}}{2m} dt \equiv -\frac{r^2}{2m}dt, \\
A^3 &=& \frac{(2m)^3 \Pi_c \Pi_s}{(\Pi_c^2 - \Pi_s^2)\Delta} dt,
\eea
and
\be
h_1 = \frac{\Delta^{\frac{1}{3}}}{2m}, \qquad h_2 = \frac{\Delta^{\frac{1}{3}}}{2m}, \qquad h_3 = 4m^2\Delta^{-\frac{2}{3}}.
\ee
These are exactly the fields given in reference \cite{Cvetic:2012tr}.

Let us now comment on the interpolating solution \cite{Baggio:2012db, Cvetic:2013cja} and the generalized notion
of subtracted geometry \cite{Chakraborty:2012nu}. From the point of view of the Harrison transformations
if instead of \eqref{groupaction:subtraction} we act with
\be
\exp [d_1 F_1 + d_2 F_2],
\ee
then we obtain an interpolating geometry, which in the limit $d_1, d_2 \to 0$ gives the original black hole and in the limit
$d_1, d_2 \to 1$ gives its subtracted geometry. If instead we exponentiate not just two negative roots but say three
\be
\exp [d_1 F_1 + d_2 F_2 + d_3 F_3],
\ee
then we obtain a bigger class of interpolating geometries, which in the limit, say $d_1 \to 1, d_2 = 0, d_3 =0$ realizes one of the
generalized subtracted geometry introduced in \cite{Chakraborty:2012nu} (upon taking the extremal limit).
Another one of these geometries can be realized as
$d_1, d_2, d_3 \to 1$.

We end this section with a discussion of rotating five-dimensional black holes.
Exactly the same transformations are applicable as in the static case. Only the expressions are more cumbersome.
The 5d Myers-Perry black hole metric in our convention is (we use $x = \cos \theta$, but otherwise our conventions are exactly the same as Cveti\v{c}-Youm \cite{Cvetic:1996xz}):
\bea
g_{tt} &=& - \frac{\Sigma - 2m}{\Sigma}, \\
g_{t\psi} &=& - \frac{2 m l_2 x^2}{\Sigma}, \\
g_{t \phi} &=& - \frac{2 m l_1 (1-x^2)}{\Sigma} \\
g_{\psi \psi} &=& \frac{x^2}{\Sigma} \left((r^2 + l_2^2)\Sigma + 2 m l_2^2 x^2\right), \\
g_{\phi \phi} & =&   \frac{1-x^2}{\Sigma} \left((r^2 + l_1^2)\Sigma + 2 m l_1^2 (1-x^2)\right), \\
g_{\psi \phi} &=& \frac{2 m l_1 l_2 x^2 (1-x^2)}{\Sigma}, \\
g_{rr} &=& \frac{\Sigma r^2}{(r^2 + l_1^2)(r^2 + l_2^2)- 2m r^2}, \\
g_{xx} &=& \frac{\Sigma}{1-x^2}.
\eea
where
\be
\Sigma = r^2 + l_1^2 x^2 + l_2^2 (1-x^2).
\ee
We do KK reduction over $t$ and $\psi$. The resulting three-dimensional metric is
\bea
ds_3^2 &=& \left[\frac{(r^2 + l_2^2)(\Sigma - 2m) + 2 m l_2^2 x^2}{r^2 (r^2 + l_1^2+ l_2^2 - 2 m) + l_1^2 l_2^2} \right] r^2 x^2 dr^2
\nonumber \\
& & + \left[(r^2 + l_2^2)(\Sigma - 2m) + 2 m l_2^2 x^2 \right] \frac{x^2}{1-x^2}dx^2 \nonumber \\
& &+  \left[ r^2 (r^2 + l_1^2+ l_2^2 - 2 m) + l_1^2 l_2^2\right] x^2 (1-x^2) d\phi^2,
\eea
and the non-zero scalars are (we also use the notation $y^I = f h^I$):
\bea
U_\rom{\scriptscriptstyle{MP}} &=&\frac{1}{2}\ln\left[\frac{\left[(r^2 + l_2^2)(\Sigma - 2m) + 2 m l_2^2 x^2\right]x^2}{\sqrt{\Sigma (\Sigma - 2m)}}\right],\\
\sigma_\rom{\scriptscriptstyle{MP}} &=& 4 l_1 l_2 m x^4 \frac{(\Sigma - m)}{\Sigma (\Sigma - 2m)},\\
\zeta^0_\rom{\scriptscriptstyle{MP}} & =& \frac{2 m l_2 x^2}{\Sigma - 2m}, \\
\tilde \zeta_{0\rom{\scriptscriptstyle{MP}}} &=& -\frac{2 m l_1 x^2}{\Sigma}, \\
y_\rom{\scriptscriptstyle{MP}} &:=& y^1 = y^2 = y^3 = \sqrt{\frac{\Sigma - 2m}{\Sigma}}.
\eea
After the action\footnote{We do not list the resulting fields after the action of  $g_\rom{charging}$ because
the five-dimensional
multi-charge black hole solution is written in detail in several references, e.g.,
\cite{Cvetic:1996xz, Giusto:2004id, G22, G25, Cvetic:2012tr}.}  of $g_\rom{charging}$, $g_\rom{subtraction}$, and $g_\rom{scaling}$ the resulting scalars in terms of the MP scalars are
\bea
U &=& U_\rom{\scriptscriptstyle{MP}}, \qquad \qquad \mbox{(as expected)}\\
\sigma &=& \sigma_\rom{\scriptscriptstyle{MP}}, \qquad  \qquad \: \mbox{(as expected)} \\
y^1 &=& \frac{y_\rom{\scriptscriptstyle{MP}}}{1-y_\rom{\scriptscriptstyle{MP}}^2}, \\
y^2 &=& \frac{y_\rom{\scriptscriptstyle{MP}}}{1-y_\rom{\scriptscriptstyle{MP}}^2}, \\
y^3 &=& \frac{y_\rom{\scriptscriptstyle{MP}}}{\Pi_c^2- \Pi_s^2 y_\rom{\scriptscriptstyle{MP}}^2}, \\
\chi^1 &=& -\frac{1}{2(1-y_\rom{\scriptscriptstyle{MP}}^2)} \left(1 + y_\rom{\scriptscriptstyle{MP}}^2 - \sqrt{\gamma_1 (\Pi_c^2 -\Pi_s^2)} (1-y_\rom{\scriptscriptstyle{MP}}^2)\right),\\
\chi^2 &=& -\frac{1}{2(1-y_\rom{\scriptscriptstyle{MP}}^2)} \left(1 + y_\rom{\scriptscriptstyle{MP}}^2 - \sqrt{\gamma_2 (\Pi_c^2 -\Pi_s^2)} (1-y_\rom{\scriptscriptstyle{MP}}^2)\right), \\
\chi^3 &=&  \frac{(1-y_\rom{\scriptscriptstyle{MP}}^2)\Pi_c \Pi_s}{(\Pi_c^2 -\Pi_s^2)(\Pi_c^2 -\Pi_s^2 y_\rom{\scriptscriptstyle{MP}}^2)},
\eea
\bea
\zeta^0 &=& \zeta^0_\rom{\scriptscriptstyle{MP}} \Pi_c + \tilde \zeta_{0\rom{\scriptscriptstyle{MP}}} \Pi_s,\\
\zeta^1 &=& \frac{1}{2}\left(\zeta^0_\rom{\scriptscriptstyle{MP}} \Pi_c \left[1- \sqrt{\gamma_1(\Pi_c^2 -\Pi_s^2)}\right] -
\tilde \zeta_{0\rom{\scriptscriptstyle{MP}}} \Pi_s \left[1+ \sqrt{\gamma_1(\Pi_c^2 -\Pi_s^2)}\right] \right),\\
\zeta^2 &=& \frac{1}{2}\left(\zeta^0_\rom{\scriptscriptstyle{MP}} \Pi_c \left[1- \sqrt{\gamma_2(\Pi_c^2 -\Pi_s^2)}\right] -
\tilde \zeta_{0\rom{\scriptscriptstyle{MP}}} \Pi_s \left[1+ \sqrt{\gamma_2(\Pi_c^2 -\Pi_s^2)}\right] \right), \\
\zeta^3 &=& -\frac{1}{\Pi_c^2 - \Pi_s^2}\left(\zeta^0_\rom{\scriptscriptstyle{MP}} \Pi_s + \tilde \zeta_{0\rom{\scriptscriptstyle{MP}}} \Pi_c\right),
\eea
\bea
\tilde \zeta_0 &=& \frac{1}{4(\Pi_c^2 - \Pi_s^2)} \Big{\{}
\zeta^0_\rom{\scriptscriptstyle{MP}} \Pi_s \left[1- \sqrt{\gamma_1(\Pi_c^2 -\Pi_s^2)}\right]\left[1- \sqrt{\gamma_2(\Pi_c^2 -\Pi_s^2)}\right] \nonumber \\
& & \qquad \quad \qquad
+ \tilde \zeta_{0\rom{\scriptscriptstyle{MP}}} \Pi_c \left[1+ \sqrt{\gamma_1(\Pi_c^2 -\Pi_s^2)}\right]\left[1+ \sqrt{\gamma_2(\Pi_c^2 -\Pi_s^2)}\right] \Big{\}},\\
%%%%%%%%%%%%%%%%%%%%%%%%%%%%%%%%%%%%%%%%%%
\tilde \zeta_1 &=& \frac{1}{2(\Pi_c^2 - \Pi_s^2)} \Big{\{}
\tilde \zeta_{0\rom{\scriptscriptstyle{MP}}} \Pi_c \left[1+ \sqrt{\gamma_2(\Pi_c^2 -\Pi_s^2)}\right]
-\zeta^0_\rom{\scriptscriptstyle{MP}} \Pi_s \left[1- \sqrt{\gamma_2(\Pi_c^2 -\Pi_s^2)}\right]\Big{\}}, \\
%%%%%%%%%%%%%%%%%%%%%%%%%%%%%%%%%%%%%%%%%%
\tilde \zeta_2 &=& \frac{1}{2(\Pi_c^2 - \Pi_s^2)} \Big{\{}
\tilde \zeta_{0\rom{\scriptscriptstyle{MP}}} \Pi_c \left[1+ \sqrt{\gamma_1(\Pi_c^2 -\Pi_s^2)}\right]
-\zeta^0_\rom{\scriptscriptstyle{MP}} \Pi_s \left[1- \sqrt{\gamma_1(\Pi_c^2 -\Pi_s^2)}\right]\Big{\}},\\
%%%%%%%%%%%%%%%%%%%%%%%%%%%%%%%%%%%%%%%%%%
\tilde \zeta_3 &=& \frac{1}{4} \Big{\{}
\tilde \zeta_{0\rom{\scriptscriptstyle{MP}}}
 \Pi_s \left[1+\sqrt{\gamma_1(\Pi_c^2 -\Pi_s^2)}\right]\left[1+\sqrt{\gamma_2(\Pi_c^2 -\Pi_s^2)}\right] \nonumber \\
& & %\qquad \quad
\quad + \zeta^0_\rom{\scriptscriptstyle{MP}} \Pi_c \left[1 - \sqrt{\gamma_1(\Pi_c^2 -\Pi_s^2)}\right]\left[1- \sqrt{\gamma_2(\Pi_c^2 -\Pi_s^2)}\right] \Big{\}}.
\eea
The five-dimensional fields constructed from the above scalars precisely match the expressions in the appendix of \cite{Cvetic:2012tr} (provided certain typos in the field $A^3$ are fixed in \cite{Cvetic:2012tr}). Explicitly, the final geometry is
\be
ds^2 = - \Delta^{-\frac{2}{3}} \Sigma(\Sigma - 2m) (dt + \cA)^2 + \Delta^{\frac{1}{3}}d\tilde s^2_4
\ee
where
\be
\Delta = (2m)^2 \Sigma (\Pi_c^2 - \Pi_s^2) + (2m)^3 \Pi_s^2,
\ee
\be
\cA = 2m (1-x^2) \left( \frac{\Pi_c}{\Sigma - 2m}l_1 - \frac{\Pi_s}{\Sigma} l_2 \right) d\phi +2m x^2 \left( \frac{\Pi_c}{\Sigma - 2m}l_2 - \frac{\Pi_s}{\Sigma} l_1 \right) d\phi,
\ee
and
\bea
d\tilde s^2_4 &=& \frac{dx^2}{1-x^2}  + \frac{r^2dr^2}{(r^2 + l_1^2)(r^2 + l_2^2) - 2mr^2} + \frac{x^2}{\Sigma} \left[r^2 + l_2^2 + \frac{2 m l_2^2 x^2}{\Sigma - 2m} \right] d\psi^2 \nn \\
& &
+ \frac{1-x^2}{\Sigma} \left[r^2 + l_1^2 + \frac{2 m l_1^2 (1- x^2)}{\Sigma - 2m} \right] d\phi^2
+ \frac{4 m l_1 l_2 x^2(1-x^2)}{\Sigma(\Sigma - 2m)} d\phi d\psi.
\eea
The scalars are
\be
h^1 = h^2 = (h^3)^{-\frac{1}{2}} = \frac{\Delta^{\frac{1}{3}}}{2m},
\ee
and the vectors are (where we have fixed minor typos in $A^3$ compared to \cite{Cvetic:2012tr}),
\be
A^1 = A^2 = - \frac{\Sigma}{2m}dt + x^2(l_1 \Pi_s - l_2 \Pi_c)d\psi + (1-x^2)(l_2 \Pi_s - l_1 \Pi_c)d\phi
\ee
\be
A^3 = \frac{(2m)^3 \Pi_s \Pi_c}{(\Pi_c^2 - \Pi_s^2)\Delta} dt
+ \frac{(2m)^3(l_1 \Pi_c - l_2 \Pi_s)}{\Delta}x^2 d\psi
+ \frac{(2m)^3(l_2 \Pi_c - l_1 \Pi_s)}{\Delta}(1-x^2) d\phi.
\ee

\subsection{Relation to Cveti\v{c}-Guica-Saleem Analysis}
Cveti\v{c}, Guica, and Saleem \cite{Cvetic:2013cja} showed that in the five-dimensional case subtracted geometry can be obtained using STU
transformations of the Euclidean STU supergravity. One reason this computation works is that both the five-dimensional
subtracted
geometry and the five-dimensional black hole have the same  4d Euclidean base space metric.
We have also obtained the same result, however, our presentation and analysis looks very different
from that of \cite{Cvetic:2013cja}. To relate the two discussions we must figure out the embedding of four-dimensional STU transformations
in the three-dimensional duality group. Having obtained this embedding if we show that the transformations we have used to obtain
subtracted geometry are from the 4d STU subgroup of the 3d duality group, then we have at least qualitatively related our analysis
to that of \cite{Cvetic:2013cja}. In fact, quantitatively too the corresponding expressions can be readily compared.

To this end we proceed along the lines of section 2 of Bossard, Nicolai, and
Stelle \cite{Bossard:2009at}. Recall that we are considering U(1)$^3$ theory in five-dimensions, which upon dimensional reduction over
a timelike direction gives rise to a Euclidean STU Einstein Maxwell theory with duality group $\mathfrak{G}_4 = \mathrm{SL}(2,\RR)^3$.
The Maxwell degrees of freedom transform
under some representation  $\mf{l}_4$ of $\mathfrak{G}_4$ (in the present case $\mathbf{8}$ of $\mathrm{SL}(2,\RR)^3$). Since we are
interested in axisymmetric configurations only, we consider them as
solutions of 3d Euclidean theory. This dimensional reduction yields one dilatonic scalar from the metric
(scalar $U$ in our notation) and one scalar each from the Maxwell field each (scalars $\zeta^\Lambda$s in our notation) together with the
scalars of the four-dimensional theory. In addition there is a vector field from the metric ($\omega_3$) and one vector field ($A_3^\Sigma$) each from the
four-dimensional Maxwell field. The Maxwell vectors become scalars after dualization in three-dimensions (scalars $\tilde \zeta_\Lambda$s in our notation).

The axisymmetric Euclidean solutions of vacuum gravity admit the so-called Ehlers symmetry $\mathrm{SL}(2,\RR)/\mathrm{SO}(2)$. This symmetry
together with the duality symmetry in four-dimensions $\mathfrak{G}_4$ results in
\be
\mathfrak{sl}(2,\RR) \oplus \mathfrak{g}_4
\ee
as a set of symmetry generators. In addition the `electric' scalars $\zeta^\Lambda$ admit shift symmetry.
After dimensional reduction the `magnetic' scalars $\tilde \zeta_\Lambda$ also admit this shift symmetry.
Since Maxwell vectors
transform in the representation $\mf{l}_4$ of $\mathfrak{G}_4$, the shift symmetries also transform in $\mf{l}_4$ of $\mathfrak{G}_4$.
%As a result in three dimensions the symmetry is enhanced to a semidirect product $\mathfrak{G}_4 \ltimes \mathfrak{l}_4$ \cite{
%Breitenlohner:1987dg, Bossard:2009we}.

The commutator of Ehlers $\mathfrak{sl}(2,\RR)$ generators on these shift symmetries give rise to new generators that also belong to
the $\mf{l}_4$ representation
of $\mathfrak{G}_4$ \cite{
Breitenlohner:1987dg, Bossard:2009at}. These new generators are also non-linearly realized on the 3d fields. Altogether, the
whole three-dimensional duality
group becomes a simple Lie group (in the present case $\mathrm{SO}(4,4)$), for which the Lie algebra
admits a five-grading with respect to the Ehlers Cartan generator:
\be
\mathfrak{g} \simeq \mathfrak{sl}(2,\RR) \oplus \mathfrak{g}_4 \oplus (2 \otimes \mathfrak{l}_4) \simeq \mathbf{1}^{(-2)} \oplus
\mathfrak{l}_4^{(-1)} \oplus (\mathbf{1} \oplus \mathfrak{g}_4)^{(0)} \oplus
\mathfrak{l}_4^{(1)} \oplus \mathbf{1}^{(2)}.
\ee
The key point to note is that all of $\mathfrak{g}_4$ has grading level 0 in this decomposition. We now show that all the generators
we use to obtain the subtracted geometry are at level 0 in this five-grading.

In our case Ehlers $\mathfrak{sl}(2,\RR)$ generators are\footnote{Recall that $U$ and $\sigma$ are with $H_0$ and $E_0$ respectively in equation \eqref{iwasawa}.} $(H_0, E_0, F_0)$, where $E_0$ and $F_0$ respectively have grading $+2$ and $-2$:
$[H_0, E_0] = 2 E_0, [H_0, F_0] = -2 F_0$. The generators
$E_{q^\Lambda}$ and $E_{p_\Lambda}$ have level $+1$ and $F_{q^\Lambda}$ and $F_{p_\Lambda}$ have level $-1$. In the construction of the subtracted
geometry starting from the charged black hole we do not use any of these generators. The three sets of generators $(H_I, E_I, F_I)$ and
the generator
$H_0$ have level 0. The
$\mathfrak{sl}(2,\RR) \oplus \mathfrak{sl}(2,\RR) \oplus \mathfrak{sl}(2,\RR)
$ generated by
$(H_I, E_I, F_I)$ is thus the four-dimensional duality group, and notice that these are the only generators that we have used in the construction of
the subtracted geometry starting from the charged black hole. Thus, although our computations are differently organized compared to \cite{Cvetic:2013cja},
 the duality symmetries that our analysis uses and the duality symmetry that their analysis uses are \emph{exactly} the same. Our
presentation and analysis has the advantage that it uses the more widely used notation of N=2 supergravity. It can be thought of as a
direct continuation of our previous work \cite{Virmani} and can be naturally
generalized to other supergravities. It also offers a different and a useful perspective on the analysis of
\cite{Cvetic:2013cja}. It can be an interesting exercise to understand in detail the relation between our Harrison transformations and the
spectral flows of \cite{Bena:2012wc, Cvetic:2013cja}. Such an analysis is beyond the scope of the present considerations, but perhaps it can be used to shed some light on the question of
interpreting the timelike Melvin twists of \cite{Bena:2012wc, Cvetic:2013cja}.

\subsection*{Acknowledgements}We thank Iosif Bena, Gary Gibbons, and Chethan Krishnan for discussions.  We thank the organizers of ``Workshop on Black Holes in Supergravity and M/Superstring Theory,'' Albert Einstein Institute, September 10-12, 2012, where certain aspects of this work were presented. We also thank seminar audience at IUCAA Pune and IISER Pune for providing useful feedback.

\appendix

\section{$\mathrm{SO}$(4,4) basis}
\label{app:so44}
Since the generators of $\mathfrak{so}$(4,4) Lie algebra feature prominently in our work and we make reference to the explicit basis we use at
various places, here we list all 28 generators in the fundamental representation $\mathbf{8}$. The basis we use is identical to the one used in \cite{Virmani}
and also in reference \cite{Bossard:2009we}. The symbol $E_{ij}$ denotes a $8 \times
8$ matrix with 1 in the $i$-th row and $j$-th column and 0 elsewhere.

\bea
H_0 = E_{33} + E_{44} - E_{77} - E_{88} && H_1 = E_{33} - E_{44} - E_{77} + E_{88}\nn \\
H_2 = E_{11} + E_{22} - E_{55} - E_{66} && H_3 = E_{11} - E_{22} - E_{55} + E_{66}
\eea
\bea
E_0 = E_{47} - E_{38} && E_1 = E_{87} - E_{34}\nn \\
E_2 = E_{25} - E_{16} && E_3 = E_{65}-E_{12}  \\
F_0 = E_{74} - E_{83} &&  F_1 = E_{78} - E_{43}\nn \\
F_2 = E_{52} - E_{61} && F_3 = E_{56} - E_{21}\\
E_{q{}_0} = E_{41} - E_{58} && E_{q{}_1} = E_{57} - E_{31} \nn \\
E_{q{}_2} = E_{46} - E_{28} && E_{q{}_3} = E_{42} - E_{68}   \\
F_{q{}_0} = E_{14} - E_{85} && F_{q{}_1} = E_{75} - E_{13} \nn \\
F_{q{}_2} = E_{64} - E_{82} && F_{q{}_3} = E_{24} - E_{86}      \\
E_{p{}^0} = E_{17} - E_{35} && E_{p{}^1} = E_{18} - E_{45}\nn  \\
E_{p{}^2} = E_{67} - E_{32} && E_{p{}^3} = E_{27} - E_{36}      \\
F_{p{}^0} = E_{71} - E_{53} && F_{p{}^1} = E_{81} - E_{54} \nn  \\
F_{p{}^2} = E_{76} - E_{23} && F_{p{}^3} = E_{72} - E_{63}.
\eea

\end{document}